\documentclass[preprint,prd,aps,showpacs,showkeys,nofootinbib]{revtex4}
\usepackage{graphicx}
\begin{document}

\title{125 GeV Higgs boson decay to a pair of muons in the $\mu\nu$SSM}

\author{Hai-Bin Zhang$^{a,b}$\footnote{hbzhang@hbu.edu.cn},
Zhi-Min Niu$^{a}$,
Ke-Sheng Sun$^{c}$,
Shu-Min Zhao$^{a,b}$,
Ying-Long Wang$^{a}$, %\footnote{hdwangyl@hbu.edu.cn},
Tai-Fu Feng$^{a,b,d}$\footnote{fengtf@hbu.edu.cn}}

\affiliation{$^a$Department of Physics, Hebei University, Baoding, 071002, China\\
$^b$Key Laboratory of High-precision Computation and Application of Quantum Field Theory of Hebei Province, Baoding, 071002, China\\
$^c$Department of Physics, Baoding University, Baoding, 071000, China\\
$^c$College of Physics, Chongqing University, Chongqing, 400044, China}

\begin{abstract}
Recently, the ATLAS and CMS Collaborations measured the 125 GeV Higgs boson decay to a pair of muons $h\rightarrow \mu \bar{\mu}$, which reported the signal strength relative to the standard model (SM) prediction is $1.2\pm0.6$ and $1.19^{+0.40+0.15}_{-0.39-0.14}$, respectively. In this work, we investigate the 125 GeV Higgs boson decay $h\rightarrow \mu \bar{\mu}$ at one-loop level in the $\mu$ from $\nu$ Supersymmetric Standard Model ($\mu\nu$SSM). Compared to the SM prediction, the decay width of $h\rightarrow \mu \bar{\mu}$ in the $\mu\nu$SSM can boost up about 10\%, considering the constraint from the muon anomalous magnetic dipole moment.
\end{abstract}

\keywords{Supersymmetry, Higgs boson decay}
\pacs{12.60.Jv, 14.80.Da}

\maketitle

\section{Introduction\label{sec1}}

Due that the Yukawa couplings of the 125 GeV Higgs boson to fermions of the first and second generation are small than that of the third generation, the Higgs boson decays to a pair of fermions of the first or second generation are challenging to measure, although the Higgs boson decays to a pair of fermions of the third generation are now measured accurately by the Large Hadron Collider (LHC). However, the ATLAS and CMS Collaborations recently measured the 125 GeV Higgs boson decay to a pair of muons $h\rightarrow \mu \bar{\mu}$, which reported the signal strength relative to the standard model (SM) prediction is $1.2\pm0.6$ with 2.0$\sigma$ \cite{ATLAS-h2u} and $1.19^{+0.40+0.15}_{-0.39-0.14}$ with 3.0$\sigma$ \cite{CMS-h2u}, respectively. The dimuon decay of the 125 GeV Higgs boson  $h\rightarrow \mu \bar{\mu}$ offers the best opportunity to measure the Higgs interactions with the second-generation fermions at the LHC. Within various theoretical frameworks, the 125 GeV Higgs boson decay $h\rightarrow \mu \bar{\mu}$ has been discussed \cite{Huu-M1,Huu-M2,Huu-M3,Huu-M4,Huu-M5,Huu-M6,Huu-M7,Huu-M8,Huu-M9,Huu-M10}. Here, we will detailedly investigate the 125 GeV Higgs boson decay $h\rightarrow \mu \bar{\mu}$ at one-loop level in the $\mu$ from $\nu$ Supersymmetric Standard Model ($\mu\nu$SSM) \cite{mnSSM,mnSSM1,mnSSM1-1,mnSSM2,mnSSM2-1,Zhang1,Zhang2}.

Through introducing three singlet right-handed neutrino superfields $\hat{\nu}_i^c$ ($i=1,2,3$), the $\mu$$\nu$SSM can solve the $\mu$ problem~\cite{m-problem} of the minimal supersymmetric standard model (MSSM)~\cite{MSSM,MSSM1,MSSM2,MSSM3,MSSM4}. The corresponding superpotential of the $\mu$$\nu$SSM is given by\ \cite{mnSSM,mnSSM1}
\begin{eqnarray}
&&W={\epsilon _{ab}}\left( {Y_{{u_{ij}}}}\hat H_u^b\hat Q_i^a\hat u_j^c + {Y_{{d_{ij}}}}\hat H_d^a\hat Q_i^b\hat d_j^c
+ {Y_{{e_{ij}}}}\hat H_d^a\hat L_i^b\hat e_j^c \right)  \nonumber\\
&&\hspace{0.95cm}
+ {\epsilon _{ab}}{Y_{{\nu _{ij}}}}\hat H_u^b\hat L_i^a\hat \nu _j^c -  {\epsilon _{ab}}{\lambda _i}\hat \nu _i^c\hat H_d^a\hat H_u^b + \frac{1}{3}{\kappa _{ijk}}\hat \nu _i^c\hat \nu _j^c\hat \nu _k^c.
\label{eq-W}
\end{eqnarray}
The summation convention is implied on repeating indices in the following. In the superpotential, the effective bilinear terms $\epsilon _{ab} \varepsilon_i \hat H_u^b\hat L_i^a$ and $\epsilon _{ab} \mu \hat H_d^a\hat H_u^b$ are generated, with $\varepsilon_i= Y_{\nu _{ij}} \left\langle {\tilde \nu _j^c} \right\rangle$ and $\mu  = {\lambda _i}\left\langle {\tilde \nu _i^c} \right\rangle$,  once the electroweak symmetry is broken. The last term generates the effective Majorana masses for neutrinos at the electroweak scale, which can help to generate three tiny neutrino masses through TeV scale seesaw mechanism
\cite{mnSSM1,neutrino-mass,neu-mass1,neu-mass2,neu-mass3,neu-mass4,neu-mass5,neu-mass6}. In the $\mu$$\nu$SSM, the gravitino or the axino can be used as the dark matter candidates~\cite{mnSSM1,mnSSM1-1,neu-mass3,DM1,DM2,DM3,DM4,DM5,DM6}. The general soft SUSY-breaking terms of the $\mu\nu$SSM, $\mathcal{L}_{soft}$, can be seen in Refs.~\cite{mnSSM1,mnSSM1-1,Zhang1}.

In the $\mu\nu$SSM, the left- and right-handed sneutrino VEVs lead to the mixing of the neutral components of the Higgs doublets with the sneutrinos producing an $8\times8$ CP-even neutral scalar mass matrix, which can be seen in Refs.~\cite{mnSSM1,mnSSM1-1,Zhang1,Zhang-MASS}. The mixing gives a rich phenomenology in the Higgs sector of  the $\mu\nu$SSM. In our previous work, the Higgs boson decay modes $h\rightarrow\gamma\gamma$, $h\rightarrow VV^*$ ($V=Z,W$), $h\rightarrow f\bar{f}$ ($f=b,\tau$), $h\rightarrow \mu\tau$, $h\rightarrow Z\gamma$, and the masses of the Higgs bosons in the $\mu\nu$SSM have been researched~\cite{Zhang-MASS,hrr,hLFV,hZr}. In this paper, we will investigate the 125 GeV Higgs boson decay channel $h\rightarrow \mu \bar{\mu}$ in the $\mu\nu$SSM to see how large new physics contributions. In Sec.~\ref{sec-h}, we give the decay width of $h\rightarrow \mu \bar{\mu}$ at one-loop level. Sec.~\ref{sec-Num} and Sec.~\ref{sec-Sum} respectively show the numerical analysis and summary.

\section{$h\rightarrow l_i \bar{l}_i$ in the $\mu\nu$SSM\label{sec-h}}

\begin{figure}
\begin{center}
\begin{minipage}[c]{0.5\textwidth}
\includegraphics[width=2.5in]{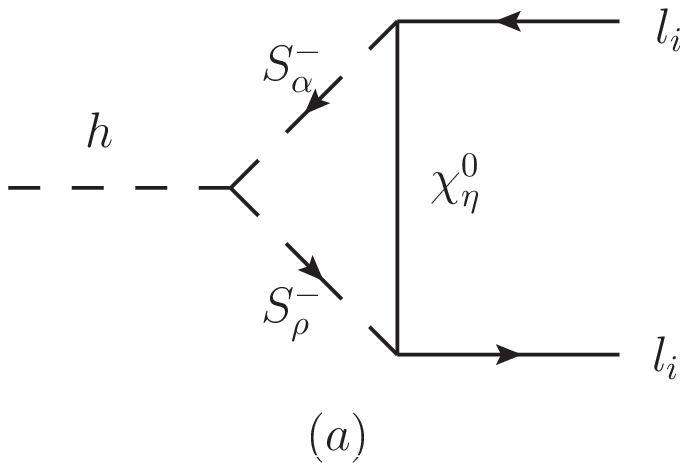}
\end{minipage}
\begin{minipage}[c]{0.39\textwidth}
\includegraphics[width=2.5in]{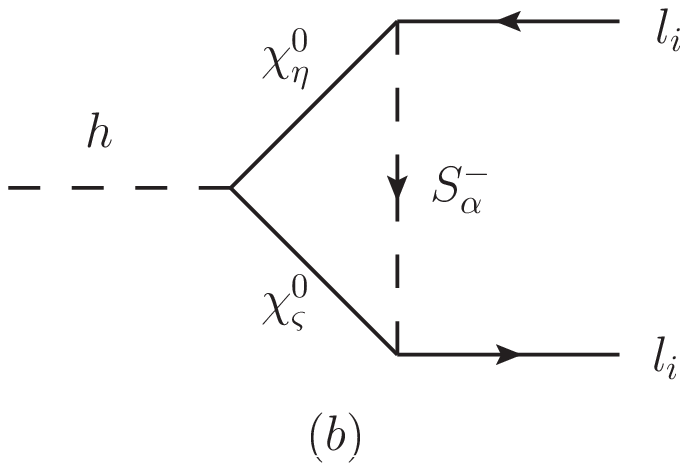}
\end{minipage}\\
\begin{minipage}[c]{0.5\textwidth}
\includegraphics[width=2.5in]{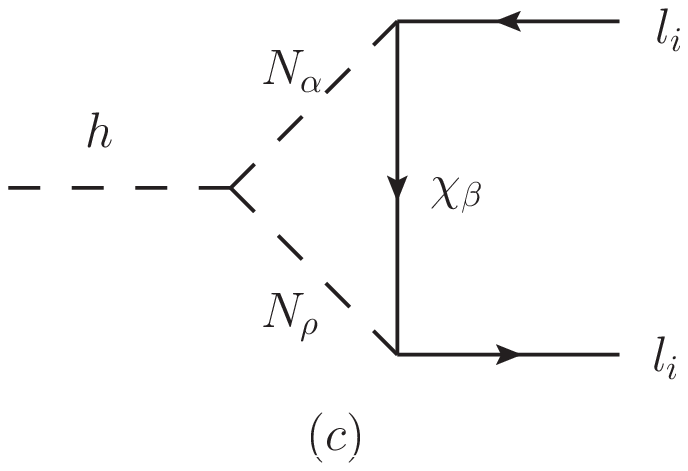}
\end{minipage}
\begin{minipage}[c]{0.39\textwidth}
\includegraphics[width=2.5in]{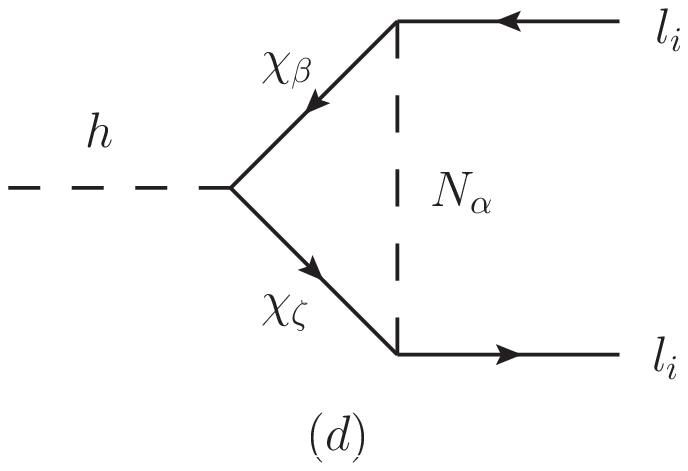}
\end{minipage}
\caption[]{The one-loop diagrams for $h\rightarrow l_i\bar{l}_i$ in the $\mu\nu$SSM. (a,b) represent the contributions from charged scalar $S_{\alpha,\rho}^-$ and neutral fermion $\chi_{\eta,\varsigma}^0$ loops, while (c,d) represent the contributions from neutral scalar $N_{\alpha,\rho}$ ($N=S,P$) and charged fermion $\chi_{\beta,\zeta}$ loops. }
\label{fig1} 
\end{center}
\end{figure}

The corresponding effective amplitude for 125 GeV Higgs decay $h\rightarrow l_i \bar{l}_i$ can be written as
\begin{eqnarray}
\mathcal{M}= {\bar l_i}({F_L^{i}}{P_L} + {F_R^{i}}{P_R}){l_i},
\end{eqnarray}
with
\begin{eqnarray}
{F_{L,R}^{i}} = F_{L,R}^{(tree)i} + F_{L,R}^{(one)i},
\end{eqnarray}
where $F_{L,R}^{(tree)i}$ denotes the contribution from the tree level, and $F_{L,R}^{(one)i}$ stands for the contribution from the one-loop level in Fig.~\ref{fig1}, respectively.

The contribution from the tree level in the $\mu\nu$SSM can be written as 
\begin{eqnarray}
F_{L}^{(tree)i}=F_{R}^{(tree)i}  = \frac{m_{l_i}}{\sqrt{2}\upsilon\cos \beta}R_{S_{11}},
\end{eqnarray}
where $m_{l_i}$  denotes the mass of the lepton $l_i$, $\upsilon\simeq174 $GeV, $R_S$ is the unitary matrix which diagonalizes the mass matrix of CP-even neutral scalars~\cite{Zhang1}. For the standard model (SM), the contribution from the tree level can be written by
\begin{eqnarray}
F_{L(\rm{SM})}^{(tree)i}=F_{R(\rm{SM})}^{(tree)i}  = \frac{m_{l_i}}{\sqrt{2}\upsilon}.
\end{eqnarray}

The one-loop diagrams for $h\rightarrow l_i\bar{l}_i$ in the $\mu\nu$SSM are depicted by Fig.~\ref{fig1}. Then, we can have
\begin{eqnarray}
F_{L,R}^{(one)i} = F_{L,R}^{(a)i} + F_{L,R}^{(b)i} + F_{L,R}^{(c)i} + F_{L,R}^{(d)i},
\end{eqnarray}
where $F_{L,R}^{(a,b)i}$ denotes the contributions from charged scalar $S_{\alpha,\rho}^-$ and neutral fermion $\chi_{\eta,\varsigma}^0$ loops, and $F_{L,R}^{(c,d)i}$ stands for the contributions from the neutral scalar $N_{\alpha,\rho}$ ($N=S,P$) and charged fermion $\chi_{\beta,\zeta}$ loops, respectively. After integrating the heavy freedoms out, we formulate the neutral fermion loop contributions $F_{L,R}^{(a,b)i}$ as follows:
\begin{eqnarray}
&&\hspace{-0.75cm}F_L^{(a)i} =  \frac{{m_{{\chi _\eta ^0}}}{C^{S^\pm}_{1 \alpha \rho }}}{{m_W^2}}
C_L^{S_\rho ^ - \chi _\eta ^0{{\bar l }_{i}}}
C_L^{S_\alpha ^{-\ast} {l_{i}}\bar \chi _\eta ^0}
{G_1}({x_{\chi _\eta ^0}},{x_{S_\alpha ^ - }},{x_{S_\rho ^ - }}) ,\nonumber\\
&&\hspace{-0.75cm}F_L^{(b)i} =  \frac{{m_{{\chi _\varsigma^0 }}}{m_{{\chi _\eta^0 }}}}{{m_W^2}}
C_L^{{S_\alpha^- }{\chi _\varsigma^0 }{{\bar l}_{i}}}C_L^{h{\chi _\eta^0 }{{\bar \chi }_\varsigma^0 }}
C_L^{{S_\alpha^{-\ast} }{l_{i}}{{\bar \chi }_\eta^0}}{G_1}({x_{{S_\alpha^- }}},{x_{{\chi _\varsigma^0 }}},{x_{{\chi _\eta^0 }}})\nonumber\\
&&\hspace{0.5cm} + \:  C_L^{{S_\alpha^- }{\chi _\varsigma^0 }{{\bar l}_{i}}}C_R^{h{\chi _\eta^0 }{{\bar \chi }_\varsigma^0 }}
C_L^{{S_\alpha^{-\ast} }{l_{i}}{{\bar \chi }_\eta^0}}{G_2}({x_{{S_\alpha^- }}},{x_{{\chi _\varsigma^0 }}},{x_{{\chi _\eta^0 }}}) ,\nonumber\\
&&\hspace{-0.75cm}F_R^{(a,b)i} = \left. {F_L^{(a,b)i}} \right|{ _{L \leftrightarrow R}} .
\end{eqnarray}
Here, the concrete expressions for couplings $C$ (and below) can be found in Ref.~\cite{hrr,hLFV}, $x= {m^2}/{m_W^2}$, $m$ is the mass for the corresponding particle, and the loop functions $G_{i}$ are given as
\begin{eqnarray}
&&{G_1}({\textit{x}_1 , \textit{x}_2 , \textit{x}_3}) =  \frac{1}{{16{\pi ^2}}}\Big[ \frac{{{x_1}\ln {x_1}}}{{({x_2} - {x_1})({x_1} - {x_3})}}
+ \frac{{{x_2}\ln {x_2}}}{{({x_1} - {x_2})({x_2} - {x_3})}} \nonumber\\
&&\hspace{2.9cm} + \frac{{{x_3}\ln {x_3}}}{{({x_1} - {x_3})({x_3} - {x_2})}}\Big], \\
&&{G_2}({\textit{x}_1 , \textit{x}_2 , \textit{x}_3}) =  \frac{1}{{16{\pi ^2}}}\Big[  \frac{{x_1^2\ln {x_1}}}{{({x_2} - {x_1})({x_1} - {x_3})}}
+ \frac{{x_2^2\ln {x_2}}}{{({x_1} - {x_2})({x_2} - {x_3})}} \nonumber\\
&&\hspace{2.9cm}  + \frac{{x_3^2\ln {x_3}}}{{({x_1} - {x_3})({x_3} - {x_2})}} \Big].\quad\;\;
\end{eqnarray}

In a similar way, the charged fermion loop contributions $F_{L,R}^{(c,d)i}$ are
\begin{eqnarray}
&&\hspace{-0.75cm}F_L^{(c)i} = \sum\limits_{N=S,P} \frac{{m_{{\chi _\beta }}}{C^{N}_{1 \alpha \rho }}}{{m_W^2}}
C_L^{N_\rho  \chi _\beta {{\bar l }_{i}}}
C_L^{N_\alpha  {l_{i}}\bar \chi _\beta }
{G_1}({x_{\chi _\beta }},{x_{N_\alpha  }},{x_{N_\rho  }}) ,\nonumber\\
&&\hspace{-0.75cm}F_L^{(d)i} = \sum\limits_{N=S,P} \Big[ C_L^{{N_\alpha }{\chi _\zeta }{{\bar l}_{i}}}C_R^{h{\chi _\beta }{{\bar \chi }_\zeta }}C_L^{{N_\alpha }{l_{i}}{{\bar \chi }_\beta }}
{G_2}({x_{{N_\alpha }}},{x_{{\chi _\zeta }}},{x_{{\chi _\beta }}})\nonumber\\
&&\hspace{-0.1cm} + \frac{{m_{{\chi _\zeta }}}{m_{{\chi _\beta }}}}{{m_W^2}}
C_L^{{N_\alpha }{\chi _\zeta }{{\bar l}_{i}}}C_L^{h{\chi _\beta }{{\bar \chi }_\zeta }}
C_L^{{N_\alpha }{l_{i}}{{\bar \chi }_\beta }}{G_1}({x_{{N_\alpha }}},{x_{{\chi _\zeta }}},{x_{{\chi _\beta }}})  \Big],\nonumber\\
&&\hspace{-0.75cm}F_R^{(c,d)i} = \left. {F_L^{(c,d)i}} \right|{ _{L \leftrightarrow R}} .
\end{eqnarray}

Then, we can obtain the decay width of $h\rightarrow l_i\bar{l}_i$
\begin{eqnarray}
{\Gamma}(h\rightarrow l_i\bar{l}_i) \simeq \frac{m_h}{16\pi}\Big({\left| {F_L^{i}} \right|^2} + {\left| {F_R^{i}} \right|^2}\Big).
\end{eqnarray}

\section{Numerical analysis\label{sec-Num}}

Firstly, we can take some appropriate parameter space in the $\mu\nu$SSM, so that we can obtain a transparent numerical results. We make the minimal flavor violation (MFV) assumptions for some parameters, which assume
\begin{eqnarray}
&&\hspace{-0.9cm}{\kappa _{ijk}} = \kappa {\delta _{ij}}{\delta _{jk}}, \quad
{({A_\kappa }\kappa )_{ijk}} =
{A_\kappa }\kappa {\delta _{ij}}{\delta _{jk}}, \quad
\lambda _i = \lambda , \nonumber\\
&&\hspace{-0.9cm}
{({A_\lambda }\lambda )}_i = {A_\lambda }\lambda,\quad
{Y_{{e_{ij}}}} = {Y_{{e_i}}}{\delta _{ij}},\quad
{({A_e}{Y_e})_{ij}} = {A_{e}}{Y_{{e_i}}}{\delta _{ij}},\nonumber\\
&&\hspace{-0.9cm}
{Y_{{\nu _{ij}}}} = {Y_{{\nu _i}}}{\delta _{ij}},\quad
(A_\nu Y_\nu)_{ij}={a_{{\nu_i}}}{\delta _{ij}},\quad
m_{\tilde \nu_{ij}^c}^2 = m_{\tilde \nu_{i}^c}^2{\delta _{ij}}, \nonumber\\
&&\hspace{-0.9cm}m_{\tilde Q_{ij}}^2 = m_{{{\tilde Q_i}}}^2{\delta _{ij}}, \quad
m_{\tilde u_{ij}^c}^2 = m_{{{\tilde u_i}^c}}^2{\delta _{ij}}, \quad
m_{\tilde d_{ij}^c}^2 = m_{{{\tilde d_i}^c}}^2{\delta _{ij}}, \nonumber\\
&&\hspace{-0.9cm}m_{{{\tilde L}_{ij}}}^2 = m_{{\tilde L}}^2{\delta _{ij}}, \quad
m_{\tilde e_{ij}^c}^2 = m_{{{\tilde e}^c}}^2{\delta _{ij}}, \quad
\upsilon_{\nu_i^c}=\upsilon_{\nu^c},
\label{MFV}
\end{eqnarray}
where $i,\;j,\;k =1,\;2,\;3 $. $m_{\tilde \nu_i^c}^2$ can be constrained by the minimization conditions of the neutral scalar potential seen in Ref.~\cite{Zhang-MASS}. To agree with experimental observations on quark mixing, one can have
\begin{eqnarray}
&&\hspace{-0.75cm}\;\,{Y_{{u _{ij}}}} = {Y_{{u _i}}}{V_{L_{ij}}^u},\quad
 (A_u Y_u)_{ij}={A_{u_i}}{Y_{{u_{ij}}}},\nonumber\\
&&\hspace{-0.75cm}\;\,{Y_{{d_{ij}}}} = {Y_{{d_i}}}{V_{L_{ij}}^d},\quad
(A_d Y_d)_{ij}={A_{d}}{Y_{{d_{ij}}}},
\end{eqnarray}
and $V=V_L^u V_L^{d\dag}$ denotes the CKM matrix.
\begin{eqnarray}
{Y_{{u_i}}} = \frac{{{m_{{u_i}}}}}{{{\upsilon_u}}},\qquad {Y_{{d_i}}} = \frac{{{m_{{d_i}}}}}{{{\upsilon_d}}},\qquad {Y_{{e_i}}} = \frac{{{m_{{l_i}}}}}{{{\upsilon_d}}},
\end{eqnarray}
where the $m_{u_{i}},m_{d_{i}}$ and $m_{l_{i}}$ stand for the up-quark, down-quark and charged lepton masses.  Through our previous work~\cite{neu-mass6}, we have discussed in detail how the neutrino oscillation data constrain neutrino Yukawa couplings $Y_{\nu_i} \sim \mathcal{O}(10^{-7})$ and left-handed sneutrino VEVs $\upsilon_{\nu_i} \sim \mathcal{O}(10^{-4}\,{\rm{GeV}})$ in the $\mu\nu$SSM via the TeV scale seesaw mechanism. Here, due to the neutrino sector only weakly affecting the decay width of $h\rightarrow \mu\bar{\mu}$, we can take no account of the constraints from neutrino experiment data.

In addition, the current difference between the experimental measurement \cite{muon-exp} and the SM theoretical prediction of the muon anomalous magnetic dipole moment (MDM) \cite{PDG1},
\begin{eqnarray}
\Delta a_{\mu}=a_{\mu}^{exp}-a_{\mu}^{SM}=(26.8\pm7.7)\times10^{-10},
\label{MDM-exp}
\end{eqnarray}
represents an discrepancy of 3.5 standard deviation, which still stands as a potential indication of the existence of new physics.
In near future, the Muon g-2 experiment E989 at Fermilab~\cite{ref-muon-exp,ref-muon-exp1} will measure the muon anomalous MDM with unprecedented precision, which may reach a 5$\sigma$ deviation from the SM, constituting an augury for new physics. In our previous work, we have studied the muon MDM at two-loop level in the $\mu\nu$SSM~\cite{hZr}. In the following, we will use the accurate theoretical prediction of the muon anomalous MDM to constrain the parameter space of the model.

Through analysis of the parameter space of the $\mu\nu$SSM in Ref.~\cite{mnSSM1}, we can take reasonable parameter values to  be $\lambda=0.1$, $\kappa=0.4$, $A_\lambda=500\;{\rm GeV}$, ${A_{\kappa}}=-300\;{\rm GeV}$ and $A_{u_{1,2}}=A_{d}=A_{e}=1\;{\rm TeV}$ for simplicity. Considering the direct search for supersymmetric particles~\cite{PDG1},  we take  $m_{{\tilde Q}_{1,2,3}}=m_{{\tilde u_{1,2}}^{c}}=m_{{\tilde d_{1,2,3}}^{c}}=2\;{\rm TeV}$, $m_{{\tilde L}}=m_{{{\tilde e}^c}}=1\;{\rm TeV}$, $M_3=2.5\;{\rm TeV}$. For simplicity, we will choose the gauginos' Majorana masses $M_1 = M_2=\mu\equiv3\lambda \upsilon_{\nu^c}$. As key parameters, $A_{u_{3}}=A_t$, $m_{{\tilde u}^c_3}$ and $\tan\beta$ greatly affect the lightest Higgs boson mass. Therefore, the free parameters that affect our next analysis are $\tan \beta ,\; \upsilon_{\nu^c},\; m_{{\tilde u}^c_3}$, and  $A_t$.

\begin{table*}
\begin{tabular*}{\textwidth}{@{\extracolsep{\fill}}lllll@{}}
\hline
Parameters&Min&Max&Step\\
\hline
$\tan \beta$&4&40&2\\
$v_{\nu^{c}}/{\rm TeV}$&1&10&0.5\\
$m_{{\tilde u}^c_3}/{\rm TeV}$&1&4&0.3\\
$A_{t}/{\rm TeV}$&1&4&0.3\\
\hline
\end{tabular*}
\caption{Scanning parameters.}
\label{tab1}
\end{table*}

To present numerical analysis, we scan the parameter space shown in Tab.~\ref{tab1}. Here the steps are large, because the running of the program is not very fast. However, the scanning parameter space is broad enough to contain the possibility of more. Considered that the light stop mass is easily ruled out by the experiment, we scan the parameter $m_{{\tilde u}^c_3}$ from 1 TeV. In the scanning, the results are constrained by the lightest Higgs boson mass in the $\mu\nu$SSM with $124.68\,{\rm GeV}\leq m_{{h}} \leq125.52\:{\rm GeV}$ \cite{PDG1}, where a $3 \sigma$ experimental error is considered. In Ref.~\cite{hrr}, we have investigated the signals of the Higgs boson decay channels $h\rightarrow\gamma\gamma$, $h\rightarrow VV^*$ ($V=Z,W$), and $h\rightarrow f\bar{f}$ ($f=b,\tau$) in the $\mu\nu$SSM. When the lightest stop mass $m_{{\tilde t}_1}\gtrsim 700\;{\rm GeV}$ and the lightest stau mass $m_{{\tilde \tau}_1}\gtrsim 300\;{\rm GeV}$, the signal strengths of these Higgs boson decay channels are in agreement with the SM. Therefore, the scanning results in this paper coincide with the experimental data of these Higgs boson decay channels.

\begin{figure}
\setlength{\unitlength}{1mm}
\centering
\begin{minipage}[c]{0.5\textwidth}
\includegraphics[width=2.8in]{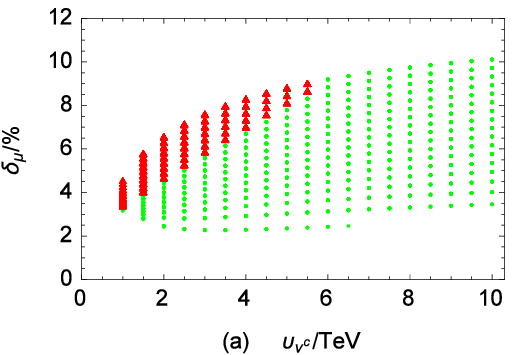}
\end{minipage}%
\begin{minipage}[c]{0.5\textwidth}
\includegraphics[width=2.85in]{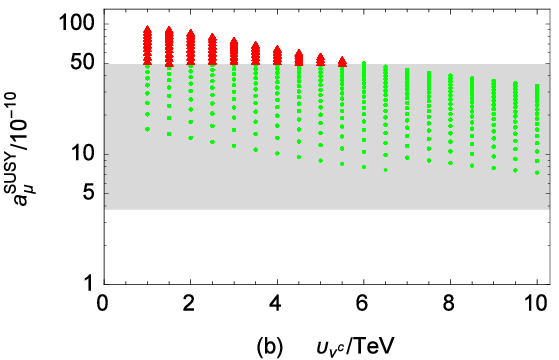}
\end{minipage}
\caption[]{The ratio $\delta_\mu$ (a) and the muon anomalous MDM $a_{\mu}^{\rm{SUSY}}$ (b)  versus the parameter $\upsilon_{\nu^{c}}$.}
\label{Fig-D2-vc}
\end{figure}

\begin{figure}
\setlength{\unitlength}{1mm}
\centering
\begin{minipage}[c]{0.5\textwidth}
\includegraphics[width=2.8in]{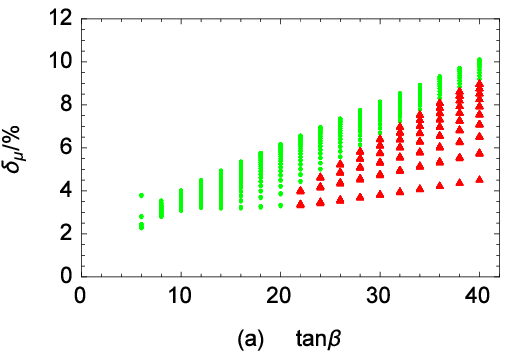}
\end{minipage}%
\begin{minipage}[c]{0.5\textwidth}
\includegraphics[width=2.85in]{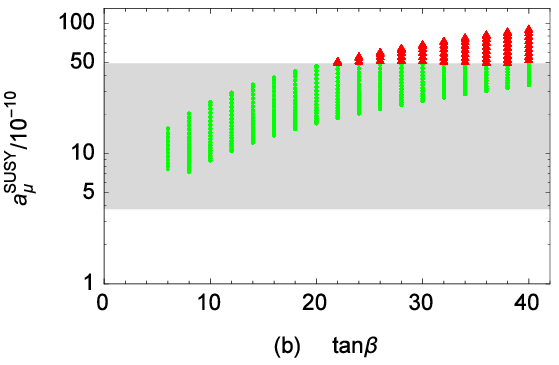}
\end{minipage}
\caption[]{The ratio $\delta_\mu$ (a) and the muon anomalous MDM $a_{\mu}^{\rm{SUSY}}$ (b)  versus the parameter  $\tan\beta$.}
\label{Fig-D2-tb}
\end{figure}

Next, we define the physical quantity
\begin{eqnarray}
\delta_\mu\equiv {{\Gamma}_{\rm{NP}}(h\rightarrow \mu\bar{\mu})-{\Gamma}_{\rm{SM}}(h\rightarrow \mu\bar{\mu})\over {\Gamma}_{\rm{SM}}(h\rightarrow \mu\bar{\mu})},
\end{eqnarray}
to show the difference of the decay width of $h\rightarrow \mu\bar{\mu}$ of the $\mu\nu$SSM (${\Gamma}_{\rm{NP}}(h\rightarrow \mu\bar{\mu})$) and that of the SM (${\Gamma}_{\rm{SM}}(h\rightarrow \mu\bar{\mu})$).
Through scanning the parameter space in Tab.~\ref{tab1}, we plot Fig.~\ref{Fig-D2-vc} and Fig.~\ref{Fig-D2-tb}, where the green dots are the corresponding physical quantity's values of the remaining parameters after being constrained by the muon anomalous MDM $a_{\mu}^{\rm{SUSY}}$ in the $\mu\nu$SSM with $3.7\times 10^{-10} \leq a_{\mu}^{\rm{SUSY}} \leq 49.9\times 10^{-10}$ considered a $3 \sigma$ experimental error. The red triangles are ruled out by the muon anomalous MDM with $a_{\mu}^{\rm{SUSY}}> 49.9\times 10^{-10}$.

In Fig.~\ref{Fig-D2-vc}, we plot the ratio $\delta_\mu$ and the muon anomalous MDM $a_{\mu}^{\rm{SUSY}}$ varying with the parameter $\upsilon_{\nu^{c}}$, where the gray area denotes
the $\Delta a_\mu$ at $3.0\sigma$ given in Eq. (\ref{MDM-exp}). In Fig.~\ref{Fig-D2-vc}(b), the numerical results show that the muon anomalous MDM $a_{\mu}^{\rm{SUSY}}$ decreases with increasing of $\upsilon_{\nu^{c}}$, which coincides with the decoupling theorem. When $\upsilon_{\nu^{c}}$ is small, $a_{\mu}^{\rm{SUSY}}$ can easily exceed the upper bound, which label the red  triangles.
In Fig.~\ref{Fig-D2-vc}(a), we can see that the ratio $\delta_\mu$ is non-decoupling with increasing  $\upsilon_{\nu^{c}}$. The maximum of the ratio $\delta_\mu$ can be around 10$\%$, when the parameter $\upsilon_{\nu^c}$ is large.
In the $\mu\nu$SSM, the parameter $\upsilon_{\nu^{c}}$ leads to the mixing of the neutral components of the Higgs doublets with the sneutrinos. The mixing affects the lightest Higgs boson mass and the Higgs couplings, which is different from the SM.

We also plot the ratio $\delta_\mu$ and the muon anomalous MDM $a_{\mu}^{\rm{SUSY}}$ varying with the parameter $\tan\beta$ in Fig.~\ref{Fig-D2-tb}. In Fig.~\ref{Fig-D2-tb}(b), the numerical results show that the muon anomalous MDM $a_{\mu}^{\rm{SUSY}}$ increases with increasing of $\tan\beta$. When $\tan\beta$ is large, $a_{\mu}^{\rm{SUSY}}$ can easily exceed the upper bound, which are easily excluded by the $\Delta a_\mu$ at $3.0\sigma$. In Fig.~\ref{Fig-D2-tb}(a), we can see that the ratio $\delta_\mu$ also increases with increasing of $\tan\beta$. The ratio $\delta_\mu$ can reach about 10$\%$, when the parameter $\tan\beta$ is large.

\section{Summary\label{sec-Sum}}

Considered that the ATLAS and CMS Collaborations measured the 125 GeV Higgs boson decay to a pair of muons $h\rightarrow \mu \bar{\mu}$ recently, we investigate the 125 GeV Higgs boson decay $h\rightarrow \mu \bar{\mu}$ at one-loop level in the $\mu$ from $\nu$ Supersymmetric Standard Model ($\mu\nu$SSM). Compared to the SM prediction, the decay width of $h\rightarrow \mu \bar{\mu}$ in the $\mu\nu$SSM can boost up about 10\%, considering the constraint from the muon anomalous magnetic dipole moment. In the future, high luminosity or high energy large colliders~\cite{ref-100pp,ref-HL,ref-CEPC,ref-ILC} will detect the Higgs boson decay $h\rightarrow \mu \bar{\mu}$ with high precision, which may see the indication of new physics.

\begin{acknowledgments}
\indent\indent
The work has been supported by the National Natural Science Foundation of China (NNSFC) with Grants No. 11705045, No. 11535002, No. 12075074, the youth top-notch talent support program of the Hebei Province,  and Midwest Universities Comprehensive Strength Promotion project.
\end{acknowledgments}

\end{document}